\documentclass[a4paper,11pt]{article}
\usepackage{pos}
\usepackage{bm}
\usepackage{amsmath}
\usepackage{nicefrac}
\usepackage[utf8]{inputenc}
\usepackage{subcaption}
\usepackage{wrapfig}
\usepackage{microtype}
\usepackage{enumitem}

\everymath{\displaystyle}
\DeclareMathAlphabet{\mathcal}{OMS}{cmsy}{m}{n}

\title{Progress in meson-meson scattering at large $N_\text{c}$}

\author*[a]{Jorge Baeza-Ballesteros}
\affiliation[a]{IFIC, CSIC-Universitat de València\\ 46980 Paterna, Spain.}
\affiliation[b]{Center for Theoretical Physics, Massachusetts Institute of Technology \\ Cambridge, MA 02139, USA.}
\emailAdd{jorge.baeza@ific.uv.es}

\author[a]{Pilar Hernández}

\emailAdd{m.pilar.hernandez@uv.es}

\author[b]{Fernando Romero-López}

\emailAdd{fromerol@mit.edu}

\abstract{We study the large $N_\text{c}$ scaling of meson-meson scattering amplitudes in a theory with $N_\text{f}=4$ degenerate quark flavors. We focus on two different scattering channels, one having the same quantum numbers as some recently found tetraquark states at LHCb. Using Lüscher's formalism, we study the $N_\text{c}$ dependence of the scattering phase shift and investigate the presence of exotic resonances in the scattering amplitude. We analyze the impact of including two-vector-meson and tetraquark-like operators to extract the finite-volume energies. }

\FullConference{The 40th International Symposium on Lattice Field Theory (Lattice 2023)\\
July 31st - August 4th, 2023\\
Fermi National Accelerator Laboratory \\
  Report number: MIT-CTP/5665}

\begin{document}
\maketitle

\section{The large $N_\text{c}$ limit of QCD}

The large $N_\text{c}$ or 't Hooft limit of QCD~\cite{tHooft:1973alw} is the limit in which the number of colors, $N_\text{c}$, is taken to infinity while keeping the number of quark flavors, $N_\text{f}$, constant. It constitutes a simplification of the theory which retains most of its non-perturbative properties, such as asymptotic freedom, confinement or spontaneous chiral symmetry breaking. This limit has proven to have predictive power in the low-energy regime and is often used by phenomenological approaches to QCD. However, different examples are known, like the $\Delta I=1/2$ rule~\cite{FUKUGITA1977237}, in which experimental results are not well reproduced by large $N_\text{c}$ predictions due to sizable subleading $N_\text{c}$ effects. While these corrections are difficult to estimate analytically, the lattice regularization provides a first-principles method to study them.

\begin{figure}[h!]
   \centering
  \begin{subfigure}[t]{0.47\linewidth}
\centering%
\includegraphics[width=1\textwidth,clip]{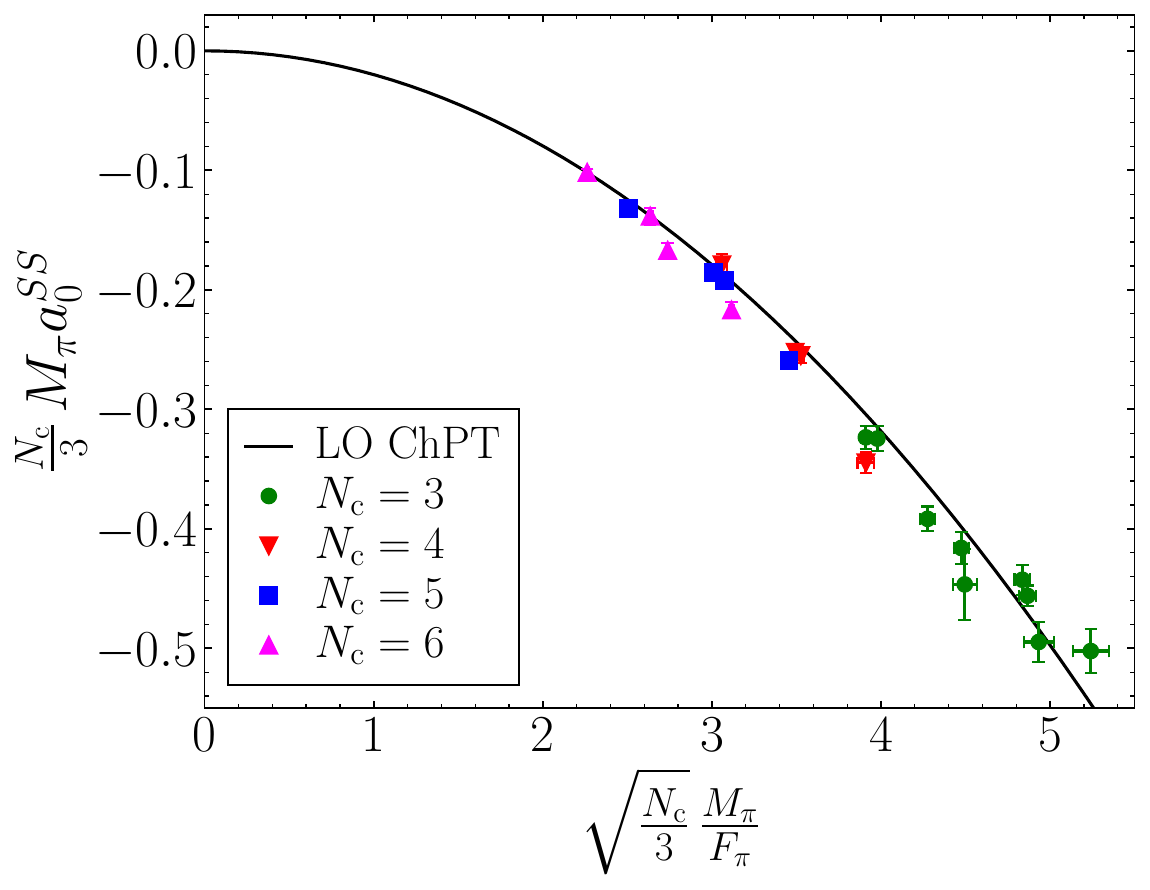}
\caption{$SS$ channel (repulsive). }
\end{subfigure}\hspace{0.01\textwidth}
   \begin{subfigure}[t]{0.47\linewidth}
\centering%
\includegraphics[width=0.967\textwidth,clip]{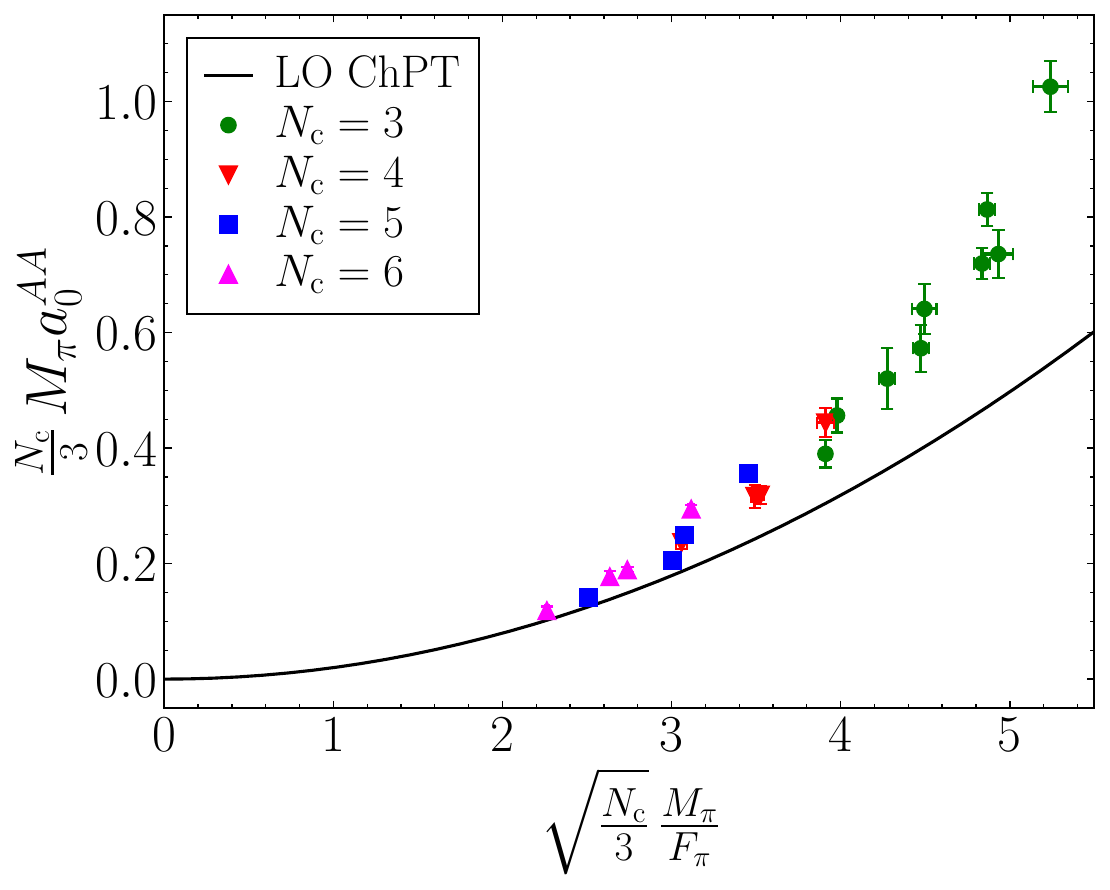}
\caption{$AA$ channel (attractive).  }
\end{subfigure}

   \caption{Results for the $s$-wave scattering length from \cite{Baeza-Ballesteros:2022azb}, together with LO ChPT predictions (black line). }
   \label{fig:scatteringlength}
\end{figure}

Several works have studied the large $N_\text{c}$ limit via lattice simulations~\cite{Hernandez:2020tbc}. In previous work, our group has investigated the scaling of different meson observables in a theory with $N_\text{f}=4$ degenerate quarks, such the pion\footnote{Since we work in a theory with $N_\text{f}=4$ degenerate quark flavors, all pseudoscalar mesons have the same mass. We refer to them generically as ``pions'' ($\pi$). Vector mesons are also degenerate and we denote them as ``rhos'' ($\rho$).} mass, $M_\pi$, and decay constant, $F_\pi$,~\cite{Hernandez:2019qed}, non-leptonic kaon decays~\cite{Donini:2020qfu}, 
 and more recently, pion-pion scattering near threshold~\cite{Baeza-Ballesteros:2022azb}. In a theory with $N_\text{f}=4$ degenerate flavors, $\pi\pi$ scattering classifies in seven scattering channels, corresponding to different irreducible representations (irreps) of the isospin symmetry group, SU(4)${}_\text{f}$~\cite{Bijnens:2011fm}. In ref.~\cite{Baeza-Ballesteros:2022azb}, we focused on two of these irreps which only contain even partial waves:
 \begin{itemize}[itemsep=0.5pt,topsep=1pt,parsep=1pt]
 \item The 84-dimensional irrep, known as the $SS$ channel, that is analogous to the isospin-2 channel of two-flavor QCD and for which a representative state is $|\pi^+\pi^+\rangle$.
 \item The 20-dimensional irrep, known as the $AA$ channel, that only exists for $N_\text{f}\geq4$ and for which a representative state is $|D_s^+\pi^+\rangle-|D^+K^+\rangle$.
 \end{itemize}
Results for the pion-pion $s$-wave scattering length, $a_0$, from ref.~\cite{Baeza-Ballesteros:2022azb} are reproduced in fig.~\ref{fig:scatteringlength}, together with leading-order (LO) predictions from chiral perturbation theory (ChPT). Note we use the sign convention for $a_0$ in which this quantity is related to the scattering phase shift as $k\cot\delta_0=1/a_0+\dots$, with  $\bm{k}$ the relative three-momentum in the center-of-mass (CM) frame and $k=|\bm{k}|$.

Interestingly, the $AA$ channel has positive $a_0$ and so it is attractive, which could lead to the presence of a tetraquark resonance at higher CM energy. This possibility is supported by various experimental findings. LHCb has recently reported several scalar tetraquark states: the $T_{cs0}^0(2900)$ in the mass spectrum of $D^-K^+$~\cite{LHCb:2020bls,LHCb:2020pxc}, and the $T_{c\overline{s}0}^{++}(2900)$ and $T_{c\overline{s}0}^0(2900)$ in the mass spectra of $D_\text{s}^+\pi^+$ and $D_\text{s}^+\pi^-$, respectively~\cite{LHCb:2022sfr,LHCb:2022lzp}. All these states have in common that, in a theory with $N_\text{f}=4$ degenerate quark flavors, they would have the quantum numbers of the $AA$ channel. LHCb has also reported a vector tetraquark in the mass spectrum of $D_s^+\pi^-$~\cite{LHCb:2022sfr,LHCb:2022lzp}, known as the $T_{c\overline{s}0}^1(2900)$, which would lie in one of the two 45-dimensional irreps of SU(4)${}_\text{f}$ in our theory,  called the $AS$ and $SA$ channels. It is worth mentioning that all these exotic states have been phenomenologically described as vector-meson molecules~\cite{Molina:2022jcd}, as they lie close to the $D^*K^*$ and $D_s^*\rho$ thresholds.

In view of these results, we are now working on an extension of ref.~\cite{Baeza-Ballesteros:2022azb} to study the scattering amplitudes of the aforementioned scattering channels as a function of the CM energy. Our objective is two-fold. First, to investigate the possible existence of tetraquark resonances and, if found, to characterize their nature as a function of $N_\text{c}$. This would allow us to answer the long-lasting question of whether these states do~\cite{Weinberg:2013cfa} or do not~\cite{coleman_1985} exist in the large $N_\text{c}$ limit, and in the former case, how their width scales with $N_\text{c}$. Second, to determine the $N_\text{c}$ dependence of meson-meson scattering amplitudes and, more specifically, to characterize the low-energy constants of ChPT as a function of $N_\text{c}$.  In this talk, we present some preliminary results for $\pi\pi$ scattering in the $SS$ channel for $N_\text{c}=3$ and $N_\text{c}=4$, and in the $AA$ channel for $N_\text{c}=3$. In the latter case, we study the impact of varying the set of operators used to determine the finite-volume energy spectrum.
 
\vspace{-0.5cm}

\section{Finite-volume energies from the lattice}

Lattice calculations allow to determine two-particle finite-volume energies, from which information about infinite-volume scattering observables can be extracted~\cite{Luscher:1986pf}. For this work, we use four ensembles with $N_\text{c}=3-6$ at fixed $M_\pi\approx590$ MeV and lattice spacing $a\approx0.075$ fm, generated with the Iwasaki gauge action and $N_\text{f}=4$ degenerate clover-improved Wilson fermions. The lattices have spatial and temporal sizes $(L/a)^3\times(T/a)=24^3\times 48$ ($N_\text{c}=3$) and $20^3\times 36$ ($N_\text{c}=4-6$). Numerical computations were performed using the HiRep code~\cite{DelDebbio:2009fd}. 

\begin{wrapfigure}[14]{r}{0.5\textwidth}
   \centering\vspace{-0.7cm}
   \begin{minipage}{0.44\textwidth}
   \includegraphics[width=1\textwidth,clip]{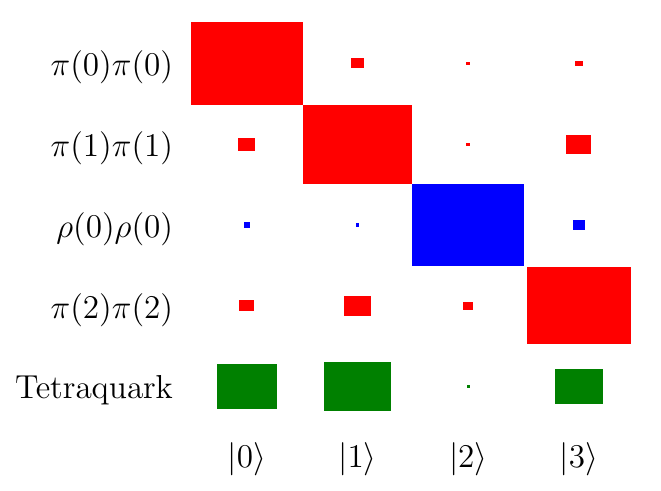}
   \caption{Relative overlap (area of the rectan-gles), of different operators into the lowest lying states of the rest-frame $A_1^+$ irrep for the $AA$ channel with $N_\text{c}=3$. Relative momentum is indicated between parenthesis in units of $2\pi/L$.}
   \label{fig:overlaps}
   \end{minipage}
\end{wrapfigure}

\vspace{0.3cm}To extract the two-particle finite-volume energies, we consider an extensive set of opera-tors and different momentum frames, which we project to the relevant irreps of the cubic group or the corresponding little group. The operator set consists of two-particle operators corresponding either to two pions, $\pi\pi$, or two vector mesons, $\rho\rho$, with various momenta assignments, as well as of local tetraquark operators. The matrix of correlators between all of these operators, $C(t)$, is determined on the lattice using time-diluted $\mathbb{Z}_2\times\mathbb{Z}_2$ noise sources for two-particle operators, and point sources located in a sparse lattice for the tetraquarks~\cite{Detmold:2019fbk}.

This matrix is then used to solve a generalized eigenvalue problem (GEVP)~\cite{Michael:1982gb,Luscher:1990ck}. The generalized eigenvectors provide insight on the overlap of the different operators into each stationary state of the finite-volume, $|n\rangle$. This is schematically shown for a few operators in fig.~\ref{fig:overlaps} in the case of the rest-frame $A_1^+$ irrep for the $AA$ channel with $N_\text{c}=3$. In the figure, the area of the rectangles is proportional to the overlap of different operators (rows) into the lowest-lying finite-volume states (columns). Two-particle operators have a dominant overlap with the state that lies the closest to the free energy associated to that particular operator---for example, in the case of the $\pi(1)\pi(1)$ operator, this would be the energy of two pions each with relative momentum $k=2\pi/L$. As for this ensemble vector mesons are stable, with mass $M_\rho\approx 1.7M_\pi$, the second excited state is dominated $\rho\rho$, while the other three states are predominantly $\pi\pi$. Regarding the tetraquark operator,  it mostly overlaps with two-pion states, with its effect on the other state being negligible.

\begin{wrapfigure}{r}{0.5\textwidth}
   \centering
   \begin{minipage}{0.48\textwidth}
   \includegraphics[width=1\textwidth,clip]{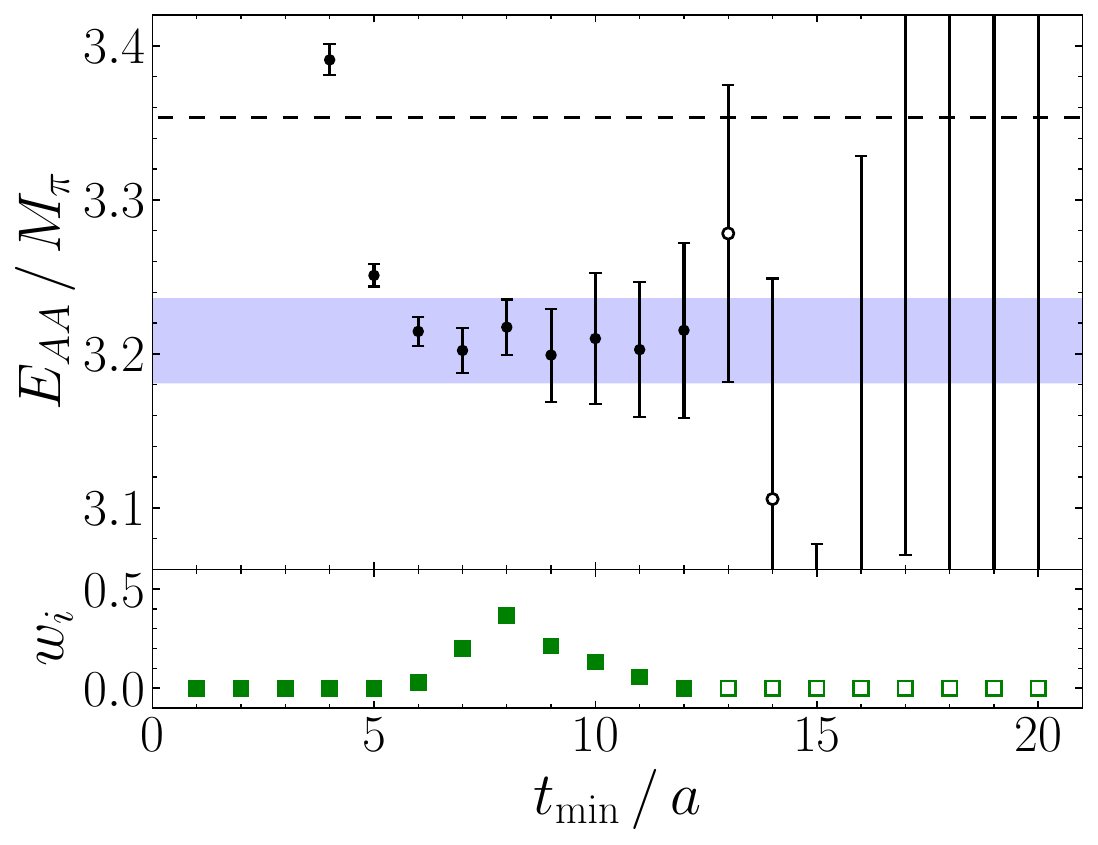}
   \caption{Best-fit results  to Eq.~\ref{eq:ratio} of the eigenvalue associated to the lowest-lying state in the [1,1,1] frame  of the $AA$ channel with $N_\text{c}=3$, for varying $t_\text{min}$. Only $\pi\pi$ and tetraquark operators are used in the GEVP. The final result is extracted averaging with weights based on the Akaike Information Criterion (lower panel). Empty points are manually excluded.  }
   \label{fig:plateaux}
   \end{minipage}
\end{wrapfigure}

 On the other hand, the generalized eigenva-lues are approximately the correlation functions between these lowest-lying  finite-volume states, and can be fitted to determine the corresponding finite-volume energies, $E_n$. Due to the limited time-extent of our ensembles, thermal effects need to be taken into account when defining the fit functions~\cite{Dudek:2012gj}. In the case of two particles with momenta $\bm{k}_1$ and $\bm{k}_2$ in the non-interacting limit, and equal mass $M$, the correlation function takes the form
 \begin{equation}
 C_{\bm{k}_1,\bm{k}_2}(t)=A\cosh\left(E_{\bm{k}_1,\bm{k}_2} \tilde{t}\right) + \tilde{A}\cosh\left(\Delta E \,\tilde{t}\right)+\dots
 \end{equation}
where $E_{\bm{k}_1,\bm{k}_2}$ is the interacting two-particle finite-volume energy, $\Delta E=E_{\bm{k}_1}-E_{\bm{k}_2}$ with $E_{\bm{k}}=\sqrt{M^2+{\bm{k}}^2}$, $A$ and $\tilde{A}$ are unknown amplitudes related to different matrix elements, $\tilde{t}=t-T/2$, and the dots refer to higher order corrections, which we neglect. Note that, while thermal effects are independent of time in the rest frame where $\bm{k}_1=\bm{k}_2$, they are in general time-dependent for moving frames, with $\bm{k}_1\neq\bm{k}_2$. Similarly, the product of two single-particle correlation functions can be decomposed as
\begin{equation}
C_{\bm{k}_1}(t)C_{\bm{k}_2}(t)=B\left[\cosh\left(E_{\bm{k}_1,\bm{k}_2}^\text{free} \tilde{t}\right) +\cosh\left(\Delta E \,\tilde{t}\right)\right]+\dots
\end{equation}
where $E_{\bm{k}_1,\bm{k}_2}^\text{free}=E_{\bm{k}_1}+E_{\bm{k}_2}$ and $B$ is another unknown amplitude.

To determine the finite-volume energies we have performed a two- or three-parameter fit, for the $\bm{k}_1=\bm{k}_2$ and $\bm{k}_1\neq\bm{k}_2$ cases, respectively,  to a ratio function
\begin{equation}\label{eq:ratio}
R(t)=\frac{\partial_0 C_{\bm{k}_1,\bm{k}_2}(t)}{\partial_0\left[C_{\bm{k}_1}(t)C_{\bm{k}_2}(t)\right]} \,
\end{equation}
for different fit ranges. The finite-volume energies are extracted where the results form a plateaux, by averaging different fit ranges using weights based on the Akaike Infomation Criteria~\cite{Jay:2020jkz}. An example of this is shown in fig.~\ref{fig:plateaux}. The top panel presents the result of the fit for different fit ranges $t\in[t_\text{min}, t_\text{max}]$ with fixed $t_\text{max}=23a$, which are averaged using the weights shown in the bottom panel to yield the blue band as a final result. Empty points are manually excluded from the average, since they are dominated by noise.

\begin{figure}[b!]
   \centering
  \begin{subfigure}[t]{0.495\linewidth}
\centering%
\includegraphics[width=1\textwidth,clip]{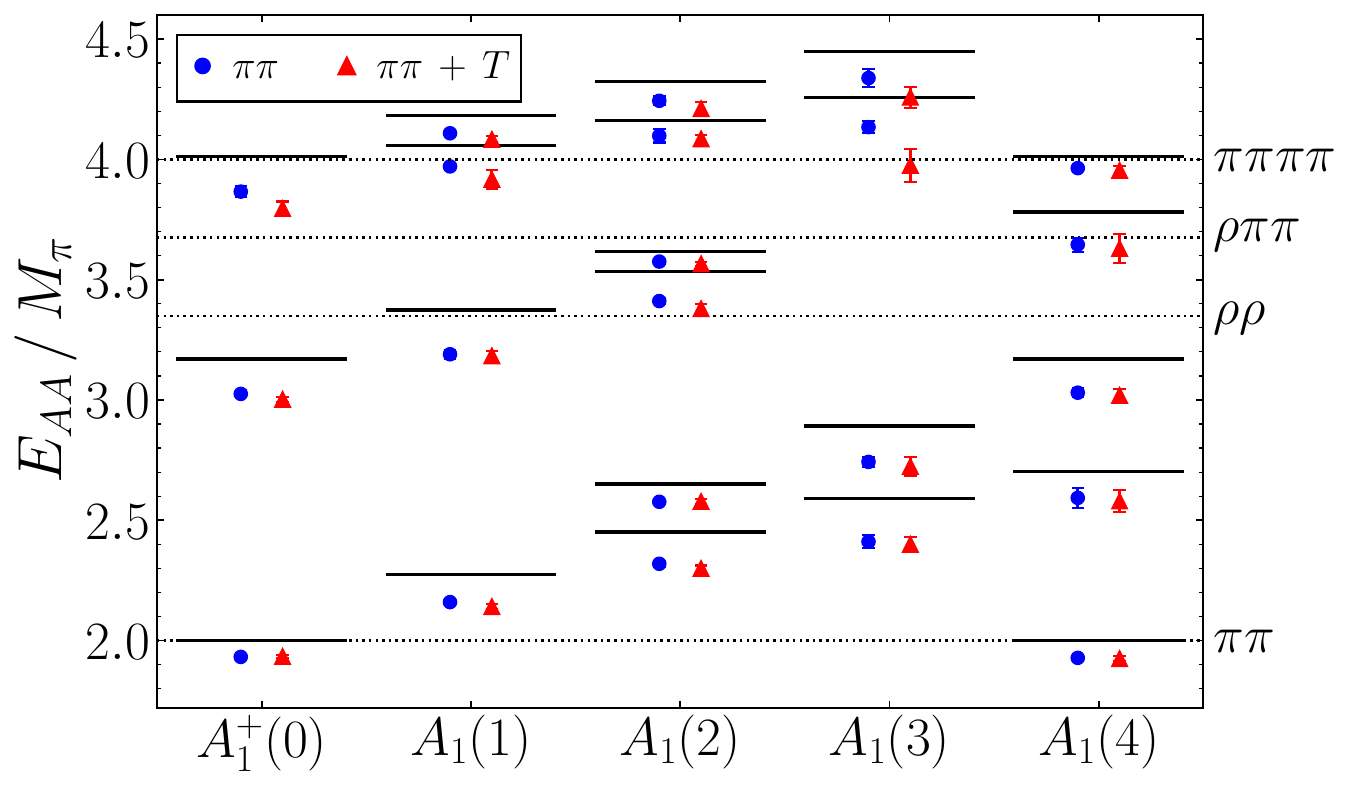}
\caption{$AA$ channel for $N_\text{c}=3$, using $\pi\pi$ (blue) and $\pi\pi$\\$+\,\text{tetraquark}$ (red) operators. }\label{fig:AAenergiespipiT}
\end{subfigure}
   \begin{subfigure}[t]{0.495\linewidth}
\centering%
\includegraphics[width=1\textwidth,clip]{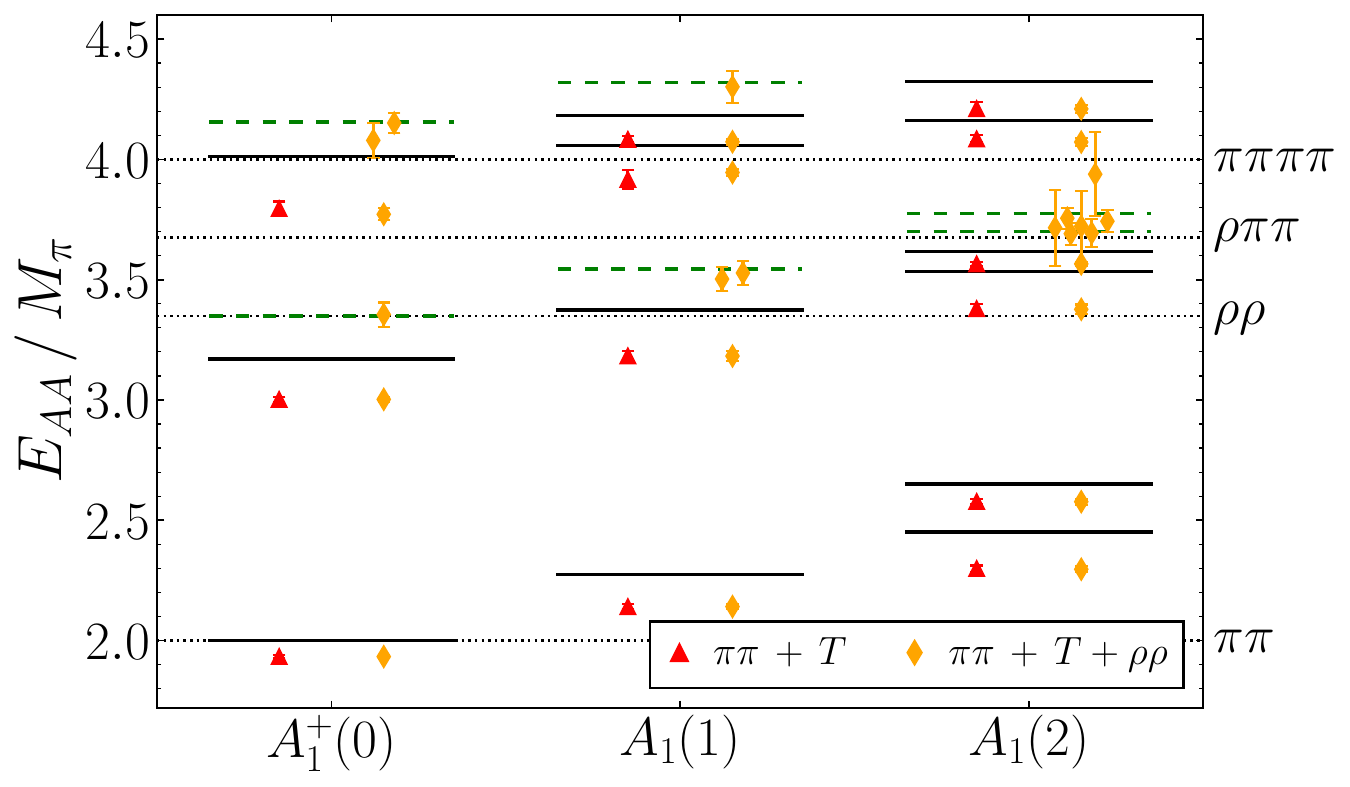}
\caption{$AA$ channel for $N_\text{c}=3$, using $\pi\pi+\text{tetraquark}$ (red) and $\pi\pi+\rho\rho+\text{tetraquark}$ (orange) operators.  }\label{fig:AAenergiespipiTrho}
\end{subfigure}\vspace{0.4cm}
\begin{subfigure}[t]{0.495\linewidth}
\centering%
\includegraphics[width=1\textwidth,clip]{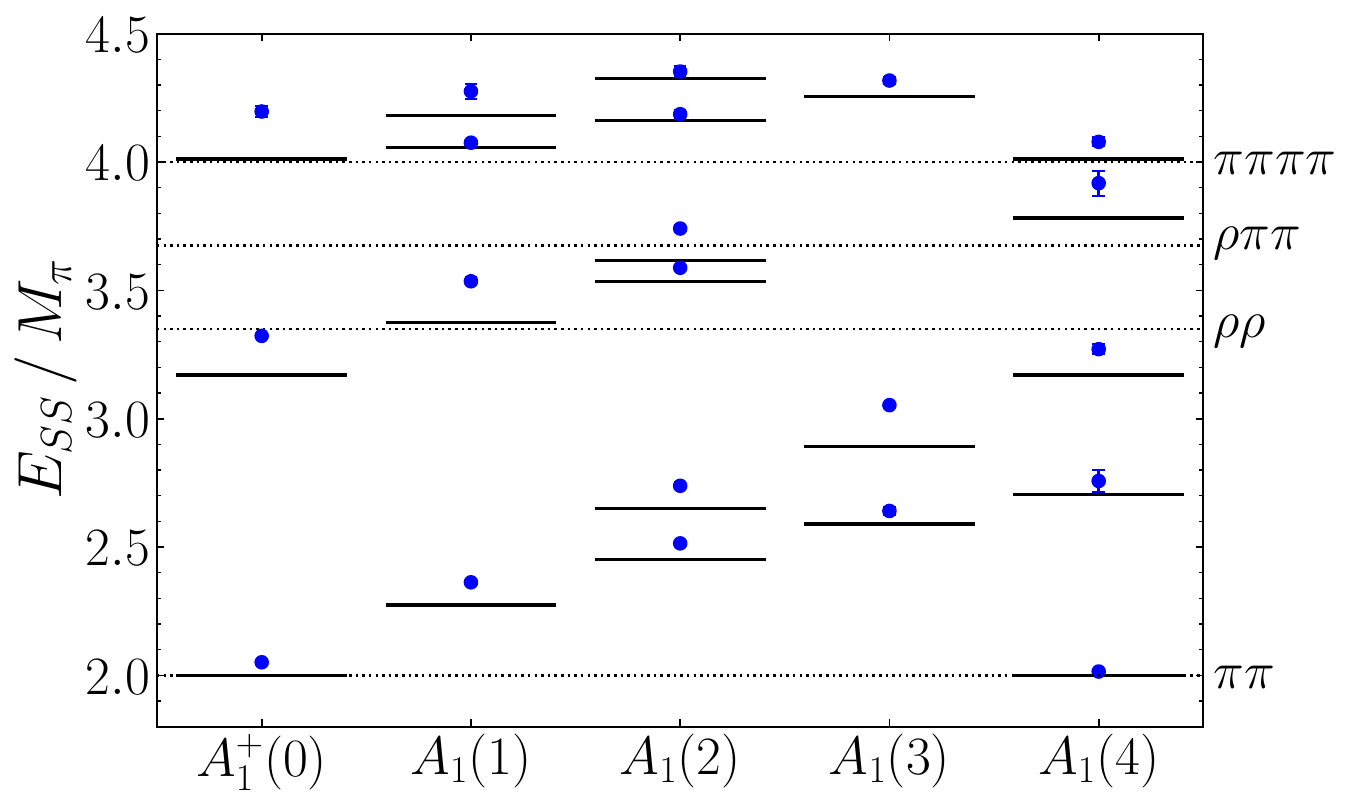}
\caption{$SS$ channel for $N_\text{c}=3$, using only $\pi\pi$ operators.  }\label{fig:SSenergies3}
\end{subfigure}
\begin{subfigure}[t]{0.495\linewidth}
\centering%
\includegraphics[width=1\textwidth,clip]{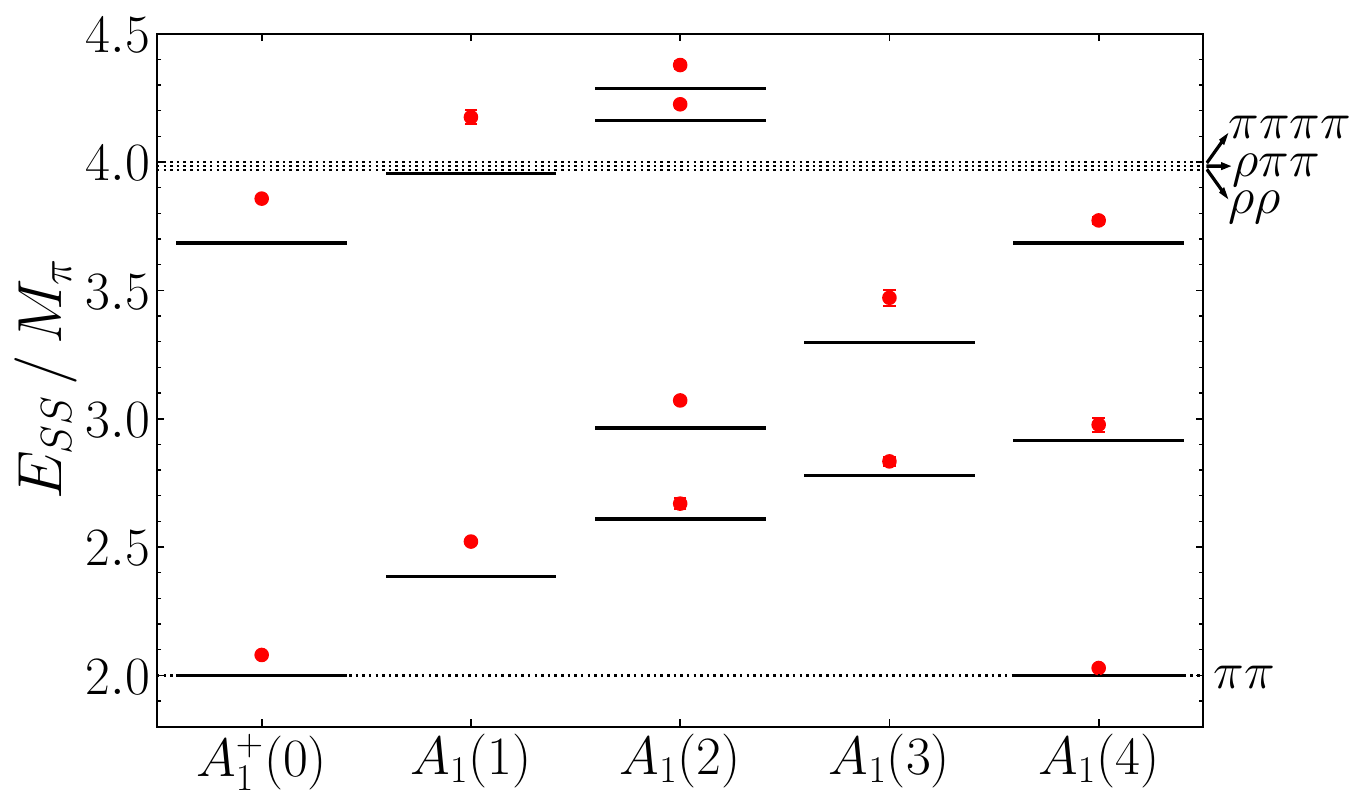}
\caption{$SS$ channel for $N_\text{c}=4$, using only $\pi\pi$ operators.  }\label{fig:SSenergies4}
\end{subfigure}\vspace{0.3cm}

   \caption{Preliminary results for finite-volume energy spectra, for different momentum frames (the magnitude of the CM momentum os shown in parenthesis in units of $2\pi/L$) and cubic-group irreps, and various choices of the operator set used to solve the GEVP. Short horizontal lines are the free energies of two pions (solid black) and two vector mesons (dashed green), and dotted lines correspond to different inelastic thresholds, as indicated next to each figure.}
   \label{fig:FVenergies}
\end{figure}

Preliminary results for the finite-volume energies are presented in fig.~\ref{fig:FVenergies} for different irreps and momentum frames. The energy spectrum of the $AA$ channel with $N_\text{c}=3$ is shown in figs.~\ref{fig:AAenergiespipiT} and~\ref{fig:AAenergiespipiTrho}, where we investigate the effect of varying the operator set used to solve the GEVP. Fig.~\ref{fig:AAenergiespipiT} compares the case in which only $\pi\pi$ operators are used against that in which tetraquark operators are also included.  Fig.~\ref{fig:AAenergiespipiTrho} compares the latter case to the result when $\rho\rho$ operators are additionally taken into account. Horizontal black and green dashed lines correspond to the $\pi\pi$ and $\rho\rho$ free energies, respectively. We note small but significant variations in the finite-volume energies when adding tetraquark operators to the operator set, especially in those energy levels close to the four-pion inelastic threshold. The inclusion of $\rho\rho$ operators, on the other hand, leads to the appearance of new energy levels associated with states of two vector mesons, while having a relatively reduced impact on those states that mainly overlap with two pions. 

In figs.~\ref{fig:SSenergies3} and~\ref{fig:SSenergies4} we present preliminary results for the finite-volume energies in the $SS$ channel for $N_\text{c}=3$ and $N_\text{c}=4$, respectively, obtained using only $\pi\pi$ operators. Note that the structure of the energy spectra is different for the two cases due to the different volumes of the two ensembles. Also, note that $M_\rho\approx1.7M_\pi$ for the $N_\text{c}=3$ ensemble, while $M_\rho\approx 2M_\pi$ for $N_\text{c}=4$.

\section{Results for the infinite-volume scattering amplitude}

The two-particle finite-volume energy spectrum can be related to infinite-volume scattering observables using the so called two-particle quantization condition. This was first proposed for two identical scalar particles in a seminal work by M. Lüscher~\cite{Luscher:1986pf}, and has since been extended to any possible two-particle process, including higher partial waves~\cite{Luscher:1990ux}, moving frames~\cite{Rummukainen:1995vs}, coupled channels~\cite{He:2005ey} and arbitrary spin~\cite{Briceno:2014oea}. In general, it takes the following form,
\begin{equation}\label{eq:QC2}
\det\left[\mathcal{K}_2^{-1}+F(L,\bm{P})\right]=0\,,
\end{equation}
where $\mathcal{K}_2$ is the infinite-volume two-particle $K$-matrix and $F$ is a geometric factor that depends on the lattice size  and the CM momentum, $\bm{P}$, and contains power-law finite-volume effects. In the case of a single-channel process dominated by the lowest partial wave, eq.~\ref{eq:QC2} can be reduced to an algebraic equation. In the case of $s$-wave, 
\begin{equation}\label{eq:QC2swave}
k\cot\delta_0=\frac{2}{\gamma L\pi^{1/2}}\mathcal{Z}^{\bm{P}}_{00}\left(\frac{kL}{2\pi}\right)\,,
\end{equation}
where $\gamma$ is the boost factor to the CM frame and $\mathcal{Z}$ is the generalized zeta function.

\begin{figure}[t!]
   \centering
  \begin{subfigure}[t]{0.48\linewidth}
\centering%
\includegraphics[width=1\textwidth,clip]{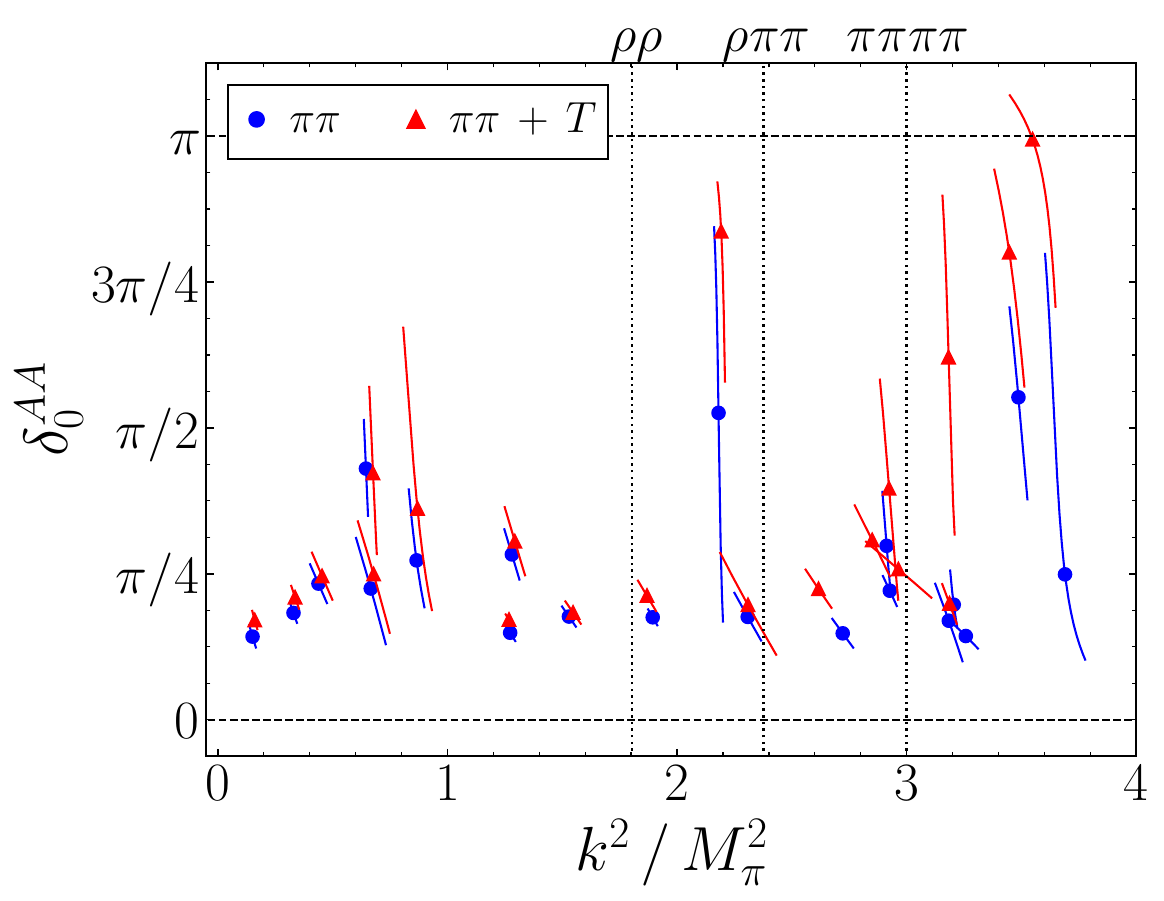}
\caption{$AA$ channel for $N_\text{c}=3$ using $\pi\pi$ (blue) and $\pi\pi+\text{tetraquark}$ (red) operators. }\label{fig:AAphaseshift}
\end{subfigure}\hspace{0.02\textwidth}
   \begin{subfigure}[t]{0.48\linewidth}
\centering%
\includegraphics[width=1\textwidth,clip]{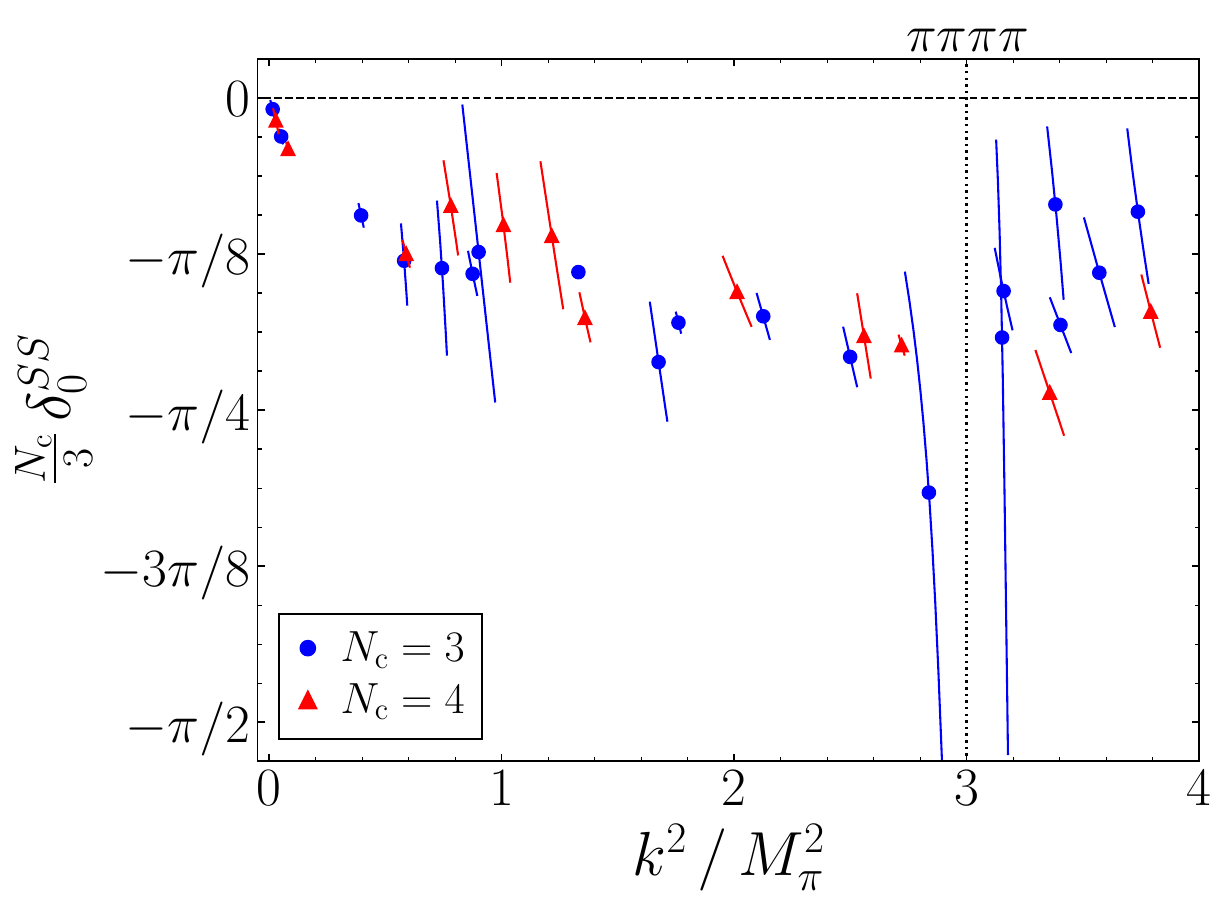}
\caption{$SS$ channel for $N_\text{c}=3$ (blue) and $N_\text{c}=4$ (red), using only $\pi\pi$ operators. For clarity, inelastic thresholds involving vector mesons are not shown. }\label{fig:SSphaseshift}
\end{subfigure}

   \caption{Preliminary results for the $s$-wave pion-pion scattering phase shift, determined using eq.~\ref{eq:QC2swave}. Dotted lines represent inelastic thresholds, as indicated over the figures. }
   
\end{figure}

Using eq.~\ref{eq:QC2swave}, we have determined the $s$-channel pion-pion scattering phase shift for the $AA$ channel using the energies levels extracted in the case only $\pi\pi$ operators are used to solve the GEVP, and also in the case they are complemented by local tetraquarks. Results are shown in fig.~\ref{fig:AAphaseshift} in blue and red, respectively. We observe both sets of operators lead to very similar phase shifts at low energies, while there is some discrepancy close to the four-pion threshold. While results for the larger operator set may suggest the presence of a resonance, note that such state would lie above the $\rho\rho$ threshold, where the application of eq.~\ref{eq:QC2swave} is just an approximation, as we are neglecting possible $\pi\pi-\rho\rho$ interactions.

Similarly, we have determined the pion-pion scattering phase shift in the $SS$ channel using the finite-volume energies shown in figs.~\ref{fig:SSenergies3} and~\ref{fig:SSenergies4}. Preliminary results for this phase shift are shown in Fig.~\ref{fig:SSphaseshift} for both $N_\text{c}=3$ and $N_\text{c}=4$, multiplied by a factor that eliminates the expected leading $N_\text{c}$ dependence. We observe the results for both $N_\text{c}$ values lie roughly on top of each other, which confirms the expected  leading $N_\text{c}$ scaling of the scattering amplitude, $\mathcal{M}\sim\mathcal{O}(N_\text{c}^{-1})$, with small subleading corrections.

\section{Summary and outlook}

In this talk, we have presented preliminary results from an ongoing study of the $N_\text{c}$ scaling of pion-pion scattering amplitudes using lattice computations, working in a theory with $N_\text{f}=4$ degenerate quark flavors. With this work, we aim at characterizing the $N_\text{c}$ dependence of the scattering amplitudes and searching for possible tetraquark resonances. So far, we have focused on the $SS$ channel for $N_\text{c}=3$ and $N_\text{c}=4$, and the $AA$ channel for $N_\text{c}=3$, with the latter having the same quantum numbers as some tetraquark states recently found at LHCb. We have determined the finite-volume energy spectra using $\pi\pi$, $\rho\rho$ and tetraquark operators, and used them to extract the corresponding pion-pion scattering phase shift. Results for the $SS$ channel follow the expected $N_\text{c}$ scaling, while those for the $AA$ channel suggest the possible presence of a tetraquark state close to the four-pion threshold. We are currently working on including higher partial waves and $\pi\pi-\rho\rho$ mixing in the quantization condition, and also on analyzing ensembles with $N_\text{c}=5$ and $N_\text{c}=6$, as well as the AS and SA channels.

\section{Acknowledgments}

JBB and PH are supported by the EU H2020 research and innovation programme under the MSC grant agreement No 860881-HIDDeN, and the Staff Exchange grant agreement No-101086085-ASYMMETRY, by the Spanish Ministerio de Ciencia e Innovaci\'on project PID2020-113644GB-I00, and by Generalitat Valenciana through the grant CIPROM/2022/69. JBB is also supported by the Spanish grant FPU19/04326 of MU. FRL is supported in part by the U.S. Department of Energy (USDOE), Office of Science, Office of Nuclear Physics, under grant Contract Numbers DE-SC0011090 and DE-SC0021006, by the Simons Foundation grant 994314 (Simons Collaboration on Confinement and QCD Strings), and by the Mauricio and Carlota Botton Fellowship. We thank Mare Nostrum 4 (BSC), Xula (CIEMAT), Finis Terrae II (CESGA), Tirant 3 (UV) and Lluis Vives (Servei d’Informàtica UV) for the computational resources provided.

\setlength{\bibsep}{0.5\itemsep}
\renewcommand\bibpreamble{\vspace{-0\baselineskip}}
\bibliographystyle{apsrev4-1.bst}
\bibliography{bibtexref.bib}

\end{document}